\documentclass[iop]{emulateapj}
\usepackage{longtable}
\usepackage{natbib}
\usepackage{multirow}
\usepackage{amsmath}
\usepackage{amsfonts}
\usepackage{graphicx}
\usepackage{color}

\newcommand{\CI}{[C\,{\sc i}]}
\newcommand{\CII}{[C\,{\sc ii}]}
\newcommand{\Cone}{[C\,{\sc i}]\,(1$-$0)}
\newcommand{\Ctwo}{[C\,{\sc i}]\,(2$-$1)}

\newcommand{\mum}{$\mu$m}

\shorttitle{\CI\ Emission in LIRGs}
\shortauthors{Jiao et al.}

\begin{document}
\bibliographystyle{apj}

\title{Neutral Carbon Emission in luminous infrared galaxies \\
The \CI\ Lines as Total Molecular Gas Tracers$^\star$}


\author{Qian Jiao\altaffilmark{1,2,3}, Yinghe Zhao\altaffilmark{4,5,6,7}, Ming Zhu\altaffilmark{1,2}, Nanyao Lu\altaffilmark{8,1},  Yu Gao\altaffilmark{7,9} and Zhi-Yu Zhang\altaffilmark{10,11}}
\altaffiltext{$\star$}{Based on Herschel observations. Herschel is an ESA space observatory with science instruments provided by European-led Principal Investigator consortia and with important participation from NASA}
\altaffiltext{1}{National Astronomical Observatories, Chinese Academy of Sciences (CAS), Beijing 100012, China; mz@nao.cas.cn}
\altaffiltext{2}{Key Laboratory of Radio Astronomy, CAS, Beijing 100012, China}
\altaffiltext{3}{University of Chinese Academy of Sciences, Beijing 100012, China}
\altaffiltext{4}{Yunnan Observatories, CAS, Kunming 650011, China; zhaoyinghe@ynao.ac.cn}
\altaffiltext{5}{Key Laboratory for the Structure and Evolution of Celestial Objects, CAS, Kunming 650011, China}
\altaffiltext{6}{Center for Astronomical Mega-Science, CAS, 20A Datun Road, Chaoyang District, Beijing 100012, China}
\altaffiltext{7}{Purple Mountain Observatory, CAS, Nanjing 210008, China}
\altaffiltext{8}{China-Chile Joint Center for Astronomy  (CCJCA), Camino El Observatorio 1515, Las Condes, Santiago, Chile; nanyao.lu@gmail.com}
\altaffiltext{9}{Key Laboratory of Radio Astronomy, CAS, Nanjing 210008, China}
\altaffiltext{10}{Institute for Astronomy, University of Edinburgh, Royal Observatory, Blackford Hill, Edinburgh EH9 3HJ, UK}
\altaffiltext{11}{ESO, Karl Schwarzschild Strasse 2, D-85748 Garching, Munich, Germany}

\begin{abstract}

\noindent We present a statistical study on the \CI\,($^{3} \rm P_{1} \rightarrow {\rm ^3 P}_{0}$), \CI\,($^{3} \rm P_{2} \rightarrow {\rm ^3 P}_{1}$) lines (hereafter \Cone\ and \Ctwo, respectively) and the CO\,(1$-$0) line for a sample of (ultra)luminous infrared galaxies [(U)LIRGs]. We explore the correlations between the luminosities of CO\,(1$-$0) and \CI\ lines, and find that $L'_\mathrm{CO(1-0)}$  correlates almost linearly with both $L'_ \mathrm{[CI](1-0)}$ and $L'_\mathrm{[CI](2-1)}$, suggesting that \CI\ lines can trace total molecular gas mass at least for (U)LIRGs. We also investigate the dependence of $L'_\mathrm{[CI](1-0)}$/$L'_\mathrm{CO(1-0)}$, $L'_\mathrm{[CI](2-1)}$/$L'_\mathrm{CO(1-0)}$ and  $L'_\mathrm{[CI](2-1)}$/$L'_\mathrm{[CI](1-0)}$ on the far-infrared color of 60-to-100\,\mum, and find non-correlation, a weak correlation and a modest correlation, respectively.  Under the assumption that these two carbon transitions are optically thin, we further calculate the  \CI\ line excitation temperatures, atomic carbon masses, and the mean \CI\ line flux-to-H$_2$ mass conversion factors for our sample. The resulting $\mathrm{H_2}$ masses using these \CI-based conversion factors roughly agree with those derived from $L'_\mathrm{CO(1-0)}$ and CO-to-H$_2$ conversion factor.
\end{abstract}

\keywords{ISM: atoms --- ISM: molecules --- galaxies: starburst --- submillimeter: galaxies}

\section{Introduction}
Carbon monoxide (CO) is the most commonly used molecular gas mass tracer in galaxies (e.g. Solomon \& Vanden Bout 2005; Bolatto et al. 2013). The low-$J$ CO transitions trace the total molecular gas mass with the CO-to-H$_2$ conversion factor ($X_{\rm CO}$). However, $X_{\rm CO}$ could vary by a factor of $\sim$10 under different physical environments (e.g., Papadopoulos et al. 2012; Bolatto et al. 2013). Furthermore, low-$J$ CO transitions in high-$z$ galaxies become difficult to observe with currently available ground facilities due to their limited sensitivities and the increasing Cosmic Microwave Background (e.g., Zhang et al. 2016). The high-$J$ CO transitions, on the other hand, can not trace the total molecular gas mass due to their high critical densities and excitation energies. Therefore, it becomes important and urgent to have some alternative H$_2$ tracers when more and more high-$z$ targets are routinely observed in high-$J$ CO transitions (e.g., Carilli \& Walter 2013). 

The emission of the two fine-structure transitions of the atomic carbon (C\,{\sc i}) in its ground state may be a particularly powerful molecular gas tracer besides canonical CO-based methods (e.g., Papadopoulos et al. 2004; Walter et al. 2011) due to the following reasons. The critical densities ($n_\mathrm{crit}$) of \CI\,($^{3} \rm P_{1} \rightarrow {\rm ^3 P}_{0}$) (rest frequency: 492.161 GHz, hereafter \Cone), and \CI\,($^{3} \rm P_{2} \rightarrow {\rm ^3 P}_{1}$) (rest frequency: 809.344 GHz, hereafter \Ctwo) are  $\sim 500\,\mathrm{cm^{-3}}$ and $\sim 10^3\, \mathrm{cm^{-3}}$ (Papadopoulos et al. 2004), respectively, which are similar to that of CO\,($J=1 \rightarrow 0$) (hereafter CO\,(1$-$0); $n_\mathrm{crit} \sim 4.4\times 10^{2} \,\mathrm{cm^{-3}}$ at kinetic temperature $T_\mathrm{kin} = 20\,$K; Yang et al. 2010).
In classical photodissociation region (PDR) models, \CI\ only exists in a narrow \CII/\CI/CO transition zone (Tielens $\&$ Hollenbach 1985). While recent observations and studies show that \CI\ can coexist with CO deep inside molecular clouds, with a remarkably constant column density ratio between \CI\ and CO (Ikeda et al. 2002). For nuclear regions or innermost disk centers of nearby galaxies, Israel (2005) found that the abundance of \CI\ is close to, or even exceeds the CO abundance. Moreover, \Cone\, and  \Ctwo\, emits $2-10$ times higher energy than CO\,(1$-$0) (even in the coldest H$_2$ gas),  so the \CI\ lines are better tracers of cold molecular gas for high-$z$ galaxies, comparing to  the observed mid/high-$J$ CO transitions which only pick up the dense and warm H$_2$ gas.  In addition, recent studies shows that cosmic rays can destroy CO (but not H$_2$) very effectively, leaving behind a C-rich phase (e.g., Bisbas et al. 2015; Krips et al. 2016), which further indicates that the optically thin \CI\ lines may represent a promising alternative to determine the total molecular gas.

Observations of \CI\ emission in nearby galaxies have been difficult due to the fact that the atmospheric transmissions at these frequencies are poor, which severely limits large surveys of the \CI\ emission in the local Universe. However the limited observations of the \CI\ emission using ground-based facilities in nearby (e.g., White et al. 1994; 
    Ojha et al. 2001; Israel 2005) and  high-$z$ (Wei$\ss$ et al. 2005, Walter et al. 2011) systems indeed suggest that \CI\ may trace H$_2$ in galaxies near and far, similarly to the low-$J$ CO lines (e.g., Zhang et al. 2014). Therefore it is important to have a large sample of local galaxies to carry out statistical studies and to compare to those of CO studies.

Thanks to the advent of the {\it Herschel Space Observatory} ({\it Herschel}; Pilbratt et al. 2010), a large number of local galaxies have been observed spectroscopically in the submillimeter window, using the {\it Herschel} Spectral and Photometric Imaging Receiver Fourier Transform Spectrometer (SPIRE/FTS; Griffin et al. 2010). Abundant ionized, atomic and molecular lines, including \Cone\ and \Ctwo, are detected (e.g., Lu et al. 2017), which allows us to carry out robust statistical analysis.

In this letter, we present  a statistical study on the \Cone\ and \Ctwo\ lines for a large sample of  local luminous infrared galaxies (LIRGs; $L_{\rm IR} \equiv L(8-1000\,\mu{\rm m}) > 10^{11}\,L_\odot$, Sanders et al. 2003) observed with the SPIRE/FTS.  The remainder of the paper is organized as follows. We briefly introduce the sample, observations, and data reduction in Section 2. The results and discussion are presented in Section 3, where we investigate the correlation between the \CI\ and CO\,(1$-$0) lines, and also calculate the conversion factors, $\alpha_\mathrm{[CI]}$ ($\equiv M({\rm {H}}_2)/L^\prime_{{\rm [CI]}}$). In the last section we summarize the main conclusions. Throughout the letter, we use a Hubble constant of $H_0 = 70\mathrm{\ km\ s^{-1}\ Mpc^{-1}}$, $\Omega_\mathrm{M}=0.3$ and $\Omega_\mathrm{\lambda}=0.7$.

\section{Sample and data reduction}

The sample discussed in this letter is selected from the program {\it A Herschel Spectroscopic Survey of Warm Molecular Gas in Local LIRGs} (PI: N. Lu), which focuses on studying the dense and warm molecular gas properties of 125 LIRGs (e.g. Lu et al. 2014, 2015). These LIRGs comprise a flux-limited subset of the Great Observatories All-Sky LIRGs Survey sample (GOALS; Armus et al. 2009). The complete dataset of the spectral lines and their fluxes for individual galaxies is given in Lu et al. (2017), and the measured \CI\ fluxes used here were based on the point-source flux calibration, as also described in Zhao et al. (2016) and Lu et al. (2014). Here we present a subsample of 71 galaxies (including 62 LIRGs and 9 ULIRGs with $L_{\rm IR} > 10^{12}\,L_\odot$), which are selected based on the following two criteria:

1. A galaxy is point-like with respect to the {\it Herschel} SPIRE beam, which is $\sim$ 35\arcsec\ at 809 GHz ( $\nu_{\rm rest} ^{\rm [CI](2-1)})$. We used the Photodetector Array Camera and Spectrometer (PACS; Poglitsch et al. 2010) 70$\,\mu$m continuum images of Chu et al. (2017) to select galaxies which are not too extended with respect to the SPIRE beam (Zhao et al. 2016, Lu et al. 2017). A galaxy is considered as a point source if its fractional 70$\,\mu$m flux within a Gaussian beam of 35\arcsec\  is $>80\%$, and we obtained 111 sources.

2. A galaxy which has CO\,(1$-$0) data (most from Sander et al. 1991; Young et al. 1995; Albrecht et al. 2007; Baan et al. 2008), with beam size greater than 35\arcsec, which results in 71 sources. For the sources which have multiple measurements, we adopted their average CO\,(1$-$0) fluxes, and used those having good signal-to-noise ratios for our calculations. The CO fluxes are shown in Table 1 and the \CI\ fluxes can be found in Lu et al. (2017). All of these 71 sources have \Ctwo\ detections, whereas only 23 have \Cone\ detections due to the reduced sensitivity near the low-frequency end of the SPIRE Long Wavelength Spectrometer Array. Typical uncertainties of the CO\,(1$-$0), \Cone\  and \Ctwo\ line fluxes are 23$\%$ (calculated from different measurements in the lierature), 13$\%$ and 8$\%$, respectively, which already include the absolute calibration uncertainty of 6$\%$ for SPIRE FTS observations (Swinyard et al. 2014).

\section{Results and Discussion}

\subsection{Relations between \CI\ and CO Emission}

\begin{figure*}[!htpb]
\centerline{\includegraphics[width=0.9\textwidth,bb=82 384 543 616]{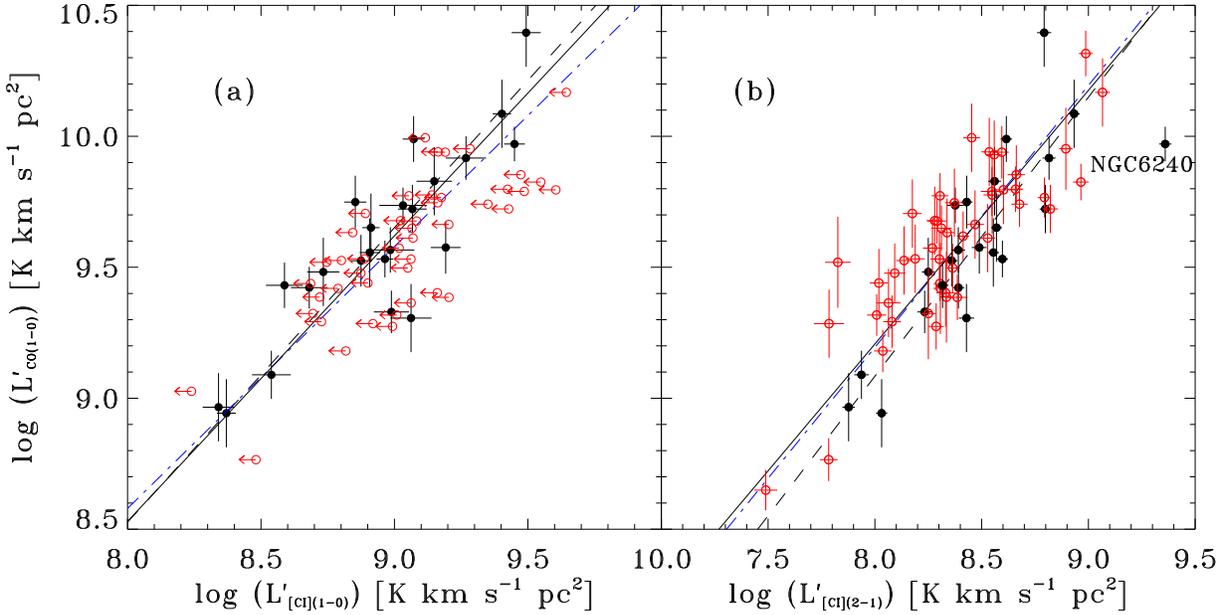}}
\caption{The luminosity of CO\,(1$-$0) is plotted against (a) \Cone\ luminosity, and (b) \Ctwo\ luminosity. The filled (black) circles represent the 23 galaxies having both \Cone\ and \Ctwo\ detections. The red (open) circles are additional galaxies with \Ctwo\ detections (panel b) and \Cone\ upper limits (panel a). For the 23 galaxies, the best-fit relations with a free slope are shown by the dashed (black) lines. Whereas the best-fit relations with a fixed slope of 1 for each detected source are shown by the dash-dot (blue) lines. The solid (black) lines show the relations for all of the 71 galaxies. NGC 6240 in panel (b) is labelled due to its largest deviation from the relation.}
\label{figure1}
\end{figure*}

Figure \ref{figure1} shows the correlations between $L'_\mathrm{CO(1-0)}$, and $L'_\mathrm{[CI](1-0)}$ (panel a) and $L'_\mathrm{[CI](2-1)}$ (panel b), which are the luminosities of CO\,(1$-$0), \Cone\ and \Ctwo, respectively. The filled (black) circles are the 23 galaxies with both \Cone\ and \Ctwo\ detected. The red (open) circles represent the other galaxies which have \Ctwo\ detections (panel b) and \Cone\ upper limits (panel a). These plots show that CO\,(1$-$0) is well correlated with both of the \CI\ lines, with the corresponding correlation coefficients of 0.81 and 0.85 in panels (a) and (b), respectively. Unweighted least-squares linear fits, using a geometrical mean functional relationship (Isobe et al. 1990), to these 23 sources with both \CI\ detections give:

\begin{equation}
\log L'_\mathrm{CO(1-0)} = (-0.47 \pm 0.76) + (1.12\pm 0.09) \ \log L'_\mathrm{[CI](1-0)},
\end{equation}

and 

\begin{equation}
\log L'_\mathrm{CO(1-0)} =  (0.56 \pm 1.55) + (1.07 \pm 0.19) \ \log L'_\mathrm{[CI](2-1)},
\end{equation}
with vertical scatters of 0.18 and 0.21 dex, respectively. The fitted trends are shown in Figure \ref{figure1} as dashed (black) lines. Furthermore, we obtain the following $L'_\mathrm{CO(1-0)}$$-$$L'_\mathrm{[CI](2-1)}$ relation for all of the 71 galaxies from a geometrical mean fitting:

\begin{equation}
\log L'_\mathrm{CO(1-0)} = (1.46 \pm 0.69) + (0.97 \pm 0.08) \ \log L'_\mathrm{[CI](2-1)},
\end{equation}
with a scatter of 0.19 dex. The fitted relation is over-plotted in Figure \ref{figure1}b with solid (black) line. 
     We also obtain the following $L'_\mathrm{CO(1-0)}$$-$$L'_\mathrm{[CI](1-0)}$ correlation (the solid black line in Figure \ref{figure1}a) using a Bayesian regression method (Kelly 2007) that also takes into consideration all the upper limits of \Cone:
 
\begin{equation}
\log L'_\mathrm{CO(1-0)} = (-0.19 \pm 1.26) + (1.09\pm 0.15) \ \log L'_\mathrm{[CI](1-0)},
\end{equation}

Equations (1)-(4) suggest that the CO\,(1$-$0) emission likely has a linear correlation with the \CI\ emission. For the \Cone\ and \Ctwo\ lines, the fitted slopes only have  marginal differences, and are consistent with each other within 1$\sigma$. The nearly linear relations between $L'_\mathrm{CO(1-0)}$ and $L'_\mathrm{[CI]}$ indicate that the CO\,(1$-$0) and \CI\ emissions might arise from similar regions within galaxies.

Considering that the low-$J$ CO emission is a commonly used tracer of the total molecular gas, the \Cone\ and \Ctwo\ lines thus can be a new avenue to determine the total molecular gas at least in (U)LIRGs. This might be particular useful for measuring the total molecular gas mass in high-z galaxies since their CO\,(1$-$0) lines are difficult to observe using ground-based facilities, whereas the \CI\ lines from distant sources become accessible for ground mm/submm telescopes. 
 
Given such a strong (and almost linear) correlation, we also fitted the $L'_\mathrm{CO(1-0)}$$-$$L'_\mathrm{[CI]}$ relations with a fixed slope of 1 in order to reduce any systemic uncertainties caused by the sample itself (e.g., sample size, dynamic range, etc.), resulting in:
\begin{equation}
	 \log L'_\mathrm{CO(1-0)}= (0.65 \pm 0.02) + \log L'_\mathrm{[CI](1-0)}, 
\end{equation}
with a scatter of 0.17 dex, and
\begin{equation}
	\log L'_\mathrm{CO(1-0)} = (1.19 \pm 0.01) + \log L'_\mathrm{[CI](2-1)},
\end{equation}
 with a scatter of 0.19 dex. The fitted relations are also plotted in Figure \ref{figure1} with dash-dot (blue) lines. If we adopt the CO conversion factor $\alpha _\mathrm{CO} \equiv M({\rm H_2})_{\rm CO}/L^\prime_{\rm {CO(1-0)}} = 0.8\,\mathrm{M_\odot}\,\mathrm {(K\,km\,s^{-1}\,pc^2)^{-1}}$ (Downes $\&$ Solomon 1998), we can obtain the \CI\ conversion factors $\alpha_\mathrm{[CI],CO}$, i.e., $\alpha_\mathrm{[CI](1-0),CO} = 3.6 \pm 0.2\,\mathrm{M_\odot}\,\mathrm {(K\,km\,s^{-1}\,pc^2)^{-1}}$
 and $\alpha_\mathrm{[CI](2-1),CO} = 12.5 \pm 0.3\,\mathrm{M_\odot}\,\mathrm {(K\,km\,s^{-1}\,pc^2)^{-1}}$, respectively. Here we only considered the fitted errors of the intercepts. A more practical choice is to take the scatters in the fitted relations as the final uncertainties, e.g., a factor of $\sim$1.5. 
 

The galaxy with the largest deviation in Figure \ref{figure1}b is NGC 6240, in which the gas heating is likely dominated by shocks (Meijerink et al. 2012). As shown in Figure 2a, NGC 6240 has the highest excitation temperature in the sample.  Therefore, the temperature influence on the level populations of C\,{\sc i} in NGC 6240 is likely the highest in this sample.  

\begin{figure}[!t]
\centerline{\includegraphics[width=0.47\textwidth,bb=107 361 416 735]{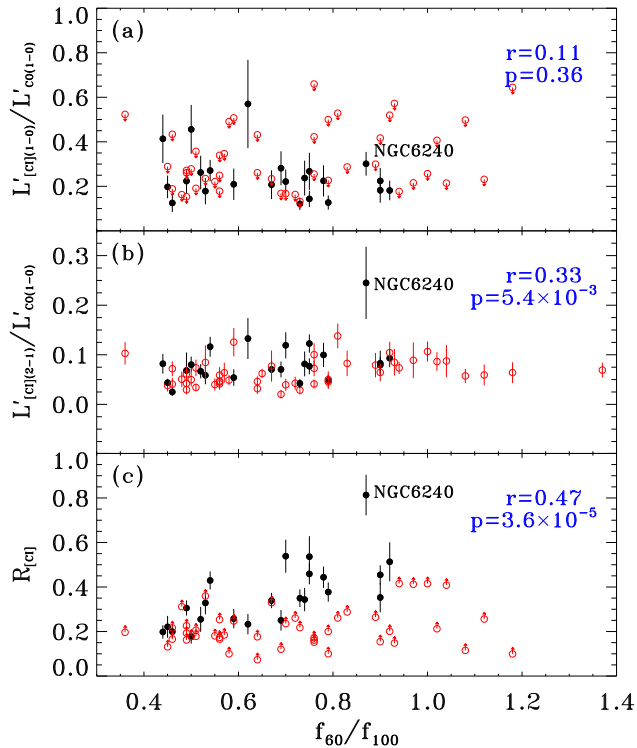}}
\caption{Plots of (a) $L'_\mathrm{[CI](1-0)}$/$L'_\mathrm{CO(1-0)}$, (b) $L'_\mathrm{[CI](2-1)}$/$L'_\mathrm{CO(1-0)}$, and (c) $R_\mathrm{[CI]}$ against  $f_{60}/f_{100}$. The open (red) circles in panels a (c) indicate the upper (lower) limits. The Labelled $r$ and $p$ represent the correlation coefficient and the possibility of no correlation respectively.}\label{figure2}
\end{figure}

In Figure \ref{figure2}, we show the luminosity ratios of $ L'_\mathrm{[CI](1-0)}$/$L'_\mathrm{CO(1-0)}$ and $L'_\mathrm{[CI](2-1)}$/$L'_\mathrm{CO(1-0)}$, as well as the \CI\ line ratio: $L'_\mathrm{[CI](2-1)}/L'_\mathrm{[CI](1-0)}$ (hereafter $R_\mathrm{[CI]}$) as a function of the IRAS $\mathrm{60\,\mu m/100\,\mu m}$ color ($f_{60}/f_{100}$) for our sample. We check possible correlations in these plots using the $cenreg$ function in NADA package within the $\bold{R}$\altaffilmark{11} statistical software environment. As labelled in panels (a) and (b) of Figures \ref{figure2}, the likelihood-$r$ coefficients are 0.11 and 0.33, respectively, with the possibilities of having no correlation at $0.36$ and $5.4\times10^{-3}$, respectively. Therefore, we conclude that there is no correlation between $L'_\mathrm{[CI](1-0)}$/$L'_\mathrm{CO(1-0)}$ and $f_{60}/f_{100}$ (Figure \ref{figure2}a). Whereas $L'_\mathrm{[CI](2-1)}$/$L'_\mathrm{CO(1-0)}$ shows a weak correlation with $f_{60}/f_{100}$ (Figure \ref{figure2}b), with a monotone increasing trend. Figure \ref{figure2}c shows that $R_\mathrm{[CI]}$ is modestly correlated with $f_{60}/f_{100}$.    
    
\footnotetext[11]{http://www.R-project.org/}
    
 The observed weak correlations involving the \Ctwo\ line might be due to the relatively higher energy required for exciting \Ctwo, comparing to that of \Cone\ and CO\,(1$-$0). \Ctwo\ is sensitive to high gas temperature, which also correlations to $f_{60}/f_{100}$ as an indicator of the intensity of the ambient UV field (thus the gas temperature, e.g. Abel et al. 2009). As shown in next subsection, the excitation temperatures of our sample galaxies are typically in the range 20$-$30\,K, which is very close to the excitation energy of \Cone\ (24$\,$K), but  significantly lower than the excitation energy of \Ctwo\ (63\,K). Thus the \Ctwo\ is more sensitive to temperature. 
    
\begin{figure*}[t]
\centerline{\includegraphics[width=0.9\textwidth,bb=46 50 250 258]{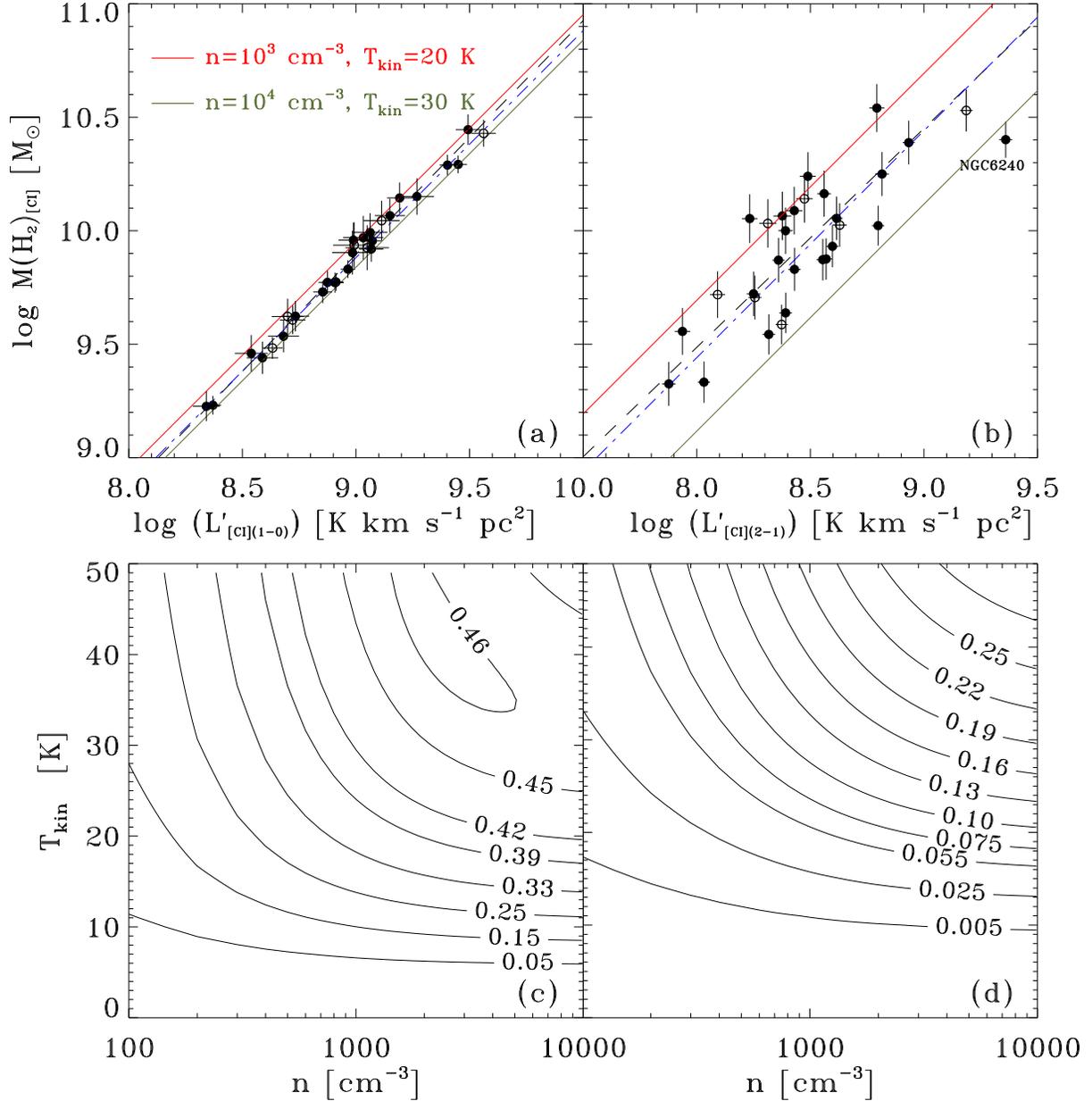}}
\caption{Panels (a) and (b) are H$_{2}$ masses calculated from the \Cone\ and \Ctwo\ emission plotted against \Cone\ and \Ctwo\ luminosities, respectively. Panels (c) and (d) show the $Q_{10}$ and $Q_{21}$ contours as a function of gas temperature ($T_\mathrm{kin}$) and density ($n$), respectively. In the top panels, the filled and open circles are galaxies with/without CO~(1$-$0) data. The best-fit  models with a fixed slope of 1 are shown by the dashed dot (blue) lines, whereas the best-fit models with a free slope are shown by the dashed (black) lines. The red and olive lines indicate the two cases with density of $10^3\,\mathrm{cm^{-3}}$ and temperature of 20$\,$K, and density of $10^4\,\mathrm{cm^{-3}}$ and temperature of 30$\,$K, respectively.}

\label{figure3}
\end{figure*}

\subsection{Total Molecular Gas Mass}

The nearly linear luminosity correlations between the \CI\ lines and CO\,(1$-$0) suggest that either \CI\ line can substitute CO\,(1$-$0) as a tracer of the  total molecular gas mass. In this subsection, we further calculate the total molecular gas mass directly from the observed \CI\ line fluxes to derive a carbon conversion factor $\alpha _\mathrm{[CI]}$, and check whether $\alpha _\mathrm{[CI]}$ agrees with $\alpha _\mathrm{[CI],CO}$. Here we also include those 7 galaxies that are detected in both \CI\ lines, but lack CO\,(1$-$0) data.

To derive the total molecular gas mass, we adopt the equation given by Papadopoulos et al. (2004) assuming optically thin \CI\ emission:
\begin{equation}
 M\mathrm{(H_2)_\mathrm{[CI]}} = \frac {\mathrm{4\pi m_{H_2}}}{hcA_{\rm ul}X_{\rm [CI]}} (\frac {{D_{\rm L}^2}}{{1+z}}) {Q_{\rm ul}^{-1}} {I_{\rm [CI]}},
\end{equation}
where ${Q_{\rm ul}= N_{\rm u}/N_{\rm [CI]}}$ is the excitation factor which depends on the gas temperature ($T_{\rm kin}$) and density (n) (Papadopoulos et al. 2004), and on the radiation field, 
$A_\mathrm{ul}$ the Einstein coefficient, with  $A_\mathrm{10}=7.93 \times 10^{-8}\, \mathrm{s^{-1}}$ and $A_\mathrm{21}=2.68 \times 10^{-7}\, \mathrm{s^{-1}}$, $X_\mathrm{[CI]}$ the abundance ratio of C\,{\sc i} to $\mathrm{H_2}$, and here we adopt $3 \times 10^{-5}$ (Wei$\ss$ et al. 2003), $z$ the redshift, and ${D_{\rm L}}$ the luminosity distance. 

The line ratio $R_\mathrm{[CI]}$ can be used to estimate the excitation temperature by adopting the equation $T_\mathrm{ex} = 38.8\, \mathrm{K/ln[2.11}/R_\mathrm{[CI]}]$ (Stutzki et al. 1997). For those 30 galaxies having both \CI\ lines detected, their $T_\mathrm{ex}$ are given in Table 2. The table shows that the excitation temperatures of most sources in our sample are in the range of $20-30$\,K. To calculate $Q_{\rm ul}$, we used equations A21 and A22 in Papadopoulos et al. (2004) to avoid assuming local thermodynamic equilibrium (LTE). These equations are derived assuming an optically thin case and using the weak radiation field approximation, i.e., $T_{\rm kin} \gg T_{\rm CMB} \sim 2.7$\,K, where $T_{\rm CMB}$ is the temperature of the cosmic microwave background. We also simply set $T_\mathrm{ex}$ as $T_{\rm kin}$ since it is difficult to have an accurate $T_{\rm kin}$ and $T_{\rm ex}$ might not deviate much from $T_{\rm kin}$ for our sample galaxies. Then at each $T_{\rm ex}$, we computed $Q_{\rm ul}$ for four gas densities of 10$^3$, 3$\times10^3$, 5$\times10^3$, $10^4\,\mathrm{cm^{-3}}$ according to Papadopoulos \& Greve (2004), and adopted the averaged value as the final $Q_{\rm ul}$. However, we find that $Q_{\rm 10}/Q_{\rm 10}^{\rm (LTE)}\sim0.94$ and $Q_{\rm 21}/Q_{\rm 21}^{\rm (LTE)}\sim0.76$, where $Q^{\rm (LTE)}_{\rm ul}$ were calculated in LTE using the same assumed $T_{\rm kin}$, indicating that our adopted physical parameters are very close to an LTE condition. The derived masses of carbon, and $\mathrm{H_2}$ are also listed in Table 2. 

In the top panels of Figure \ref{figure3}, we plot the resulting $M(\mathrm{H_2})$ as a function of $L'_\mathrm{[CI]}$.  The red and olive lines in each plot show the two cases with $(n, T_{\mathrm{kin}}) = (10^3\,\mathrm{cm^{-3}}$,  20$\,$K) and  ($10^4\,\mathrm{cm^{-3}}$, 30$\,$K), respectively. The dashed (black) lines are unweighted least-squares fits to these data points, which give:
\begin{equation}
\log M(\mathrm{H_2})_\mathrm{[CI](1-0)}= (0.63 \pm 0.19) +(1.02 \pm 0.02)\ \log L'_\mathrm{[CI](1-0)},
\end{equation}
and 
\begin{equation}
 \log M(\mathrm{H_2})_\mathrm{[CI](2-1)}= (1.78 \pm 1.00) + (0.96  \pm 0.13) \ \log L'_\mathrm{[CI](2-1)},
\end{equation}
with scatters of 0.04 dex and 0.24 dex, respectively. We also fit these models using a fixed slope of 1, and obtained:
\begin{equation}
\log M(\mathrm{H_2})_\mathrm{[CI](1-0)}= (0.88 \pm 0.01) + \log L'_\mathrm{[CI](1-0)},
\end{equation}
and 
\begin{equation}
 \log M(\mathrm{H_2})_\mathrm{[CI](2-1)}= (1.44 \pm 0.02) +  \log L'_\mathrm{[CI](2-1)},
\end{equation}
with scatters of 0.05 dex and 0.24 dex. The fitted results are plotted in Figure \ref{figure3} as dash-dotted (blue) lines. 

 As shown in Figures \ref{figure3}a and b, the derived model of $M(\mathrm{H_2})_\mathrm{[CI](1-0)}$ from $L'_\mathrm{[CI](1-0)}$ shows a smaller dispersion than that of $M(\mathrm{H_2})_\mathrm{[CI](2-1)}$ from $L'_\mathrm{[CI](2-1)}$. This is because $Q_{\rm 21}$ is much more sensitive to temperature and density than $Q_{\rm 10}$, which is illustrated clearly in Figures \ref{figure3}c and d. Hence, the \Cone\ emission is a better total molecular gas mass tracer compared to the \Ctwo\ line in the theoretical aspect if we do not have constrains on the gas temperature and density. However, in practice the variation of the C\,{\sc i} abundance (see below) and/or violation of the assumption of optically thin \CI\ emission may wash out this effect and result in similar scatter for both lines as indicated in Figure \ref{figure1}.


We calculated the conversion factors using equations (10) and (11), and obtained $\alpha_\mathrm{[CI](1-0)} = 7.6 \pm 0.2$\,$\mathrm {(K\,km\,s^{-1}\,pc^2)^{-1}}$ and $\alpha_\mathrm{[CI](2-1)} = 27.5 \pm 1.3$\,$\mathrm {(K\,km\,s^{-1}\,pc^2)^{-1}}$. These values are about 2 times larger than those (see \S3.1) derived from the \CI$-$CO relations with $\alpha_{\rm CO}=0.8\,\mathrm {(K\,km\,s^{-1}\,pc^2)^{-1}}$. Several reasons could cause these discrepancies: (1) the adopted $\alpha _\mathrm{CO}$ is smaller than its real value; (2) our adopted $X_\mathrm{[CI]}$ is too high; and/or (3) the density we adopted to calculate $Q_{\rm ul}$ is too low. Indeed, Papadopoulos et al. (2012) and Scoville et al. (2016) found that even in (U)LIRGs $\alpha_{\rm CO}$ can be as high as the Galactic value. Regarding $X_\mathrm{[CI]}$, several studies on local and high-redshift star-forming environments have shown that it is in the range of $2-16\times10^{-5}$ (
    Pety et al. 2004; Wei$\ss$ et al. 2005; Walter et al. 2011). Further, $Q_{\rm ul}$ varies by a factor of $1-2$ when the density changing from 10$^3$ to 10$^4$\,cm$^{-3}$ within our temperature range. Therefore, all of these uncertainties make our estimators have accuracies of a factor of $2-3$.

\begin{table*}
\centering
\begin{minipage}[]{120mm}
\caption{The CO\,(1$-$0) flux}\centering
\label{tab1}\end{minipage}
\setlength{\tabcolsep}{1pt}
\small
 \begin{tabular}{ccc|ccc}
  \hline\noalign{\smallskip}
   galaxy $^{a}$ &   $I_\mathrm{CO\,(1-0)}$         & Ref.$^{b}$ &  galaxy &   $I_\mathrm{CO\,(1-0)}$          & Ref.$^{b}$\\
              &     Jy km\,$\mathrm{s^{-1}}$     &    &    &    Jy km\,$\mathrm{s^{-1}}$ & \\
  \hline\noalign{\smallskip}
 
Arp 193 *                   &  182.6   $\pm$ 38.8        &      B08, P12                              &     ESO 069-IG006        & 197.1  $\pm$  39.4    &   M90\\
Arp 220 *                   &  445.3   $\pm$ 85.3       &      Y95, B08, P12                      &     ESO 148-IG002        &  59.4   $\pm$ 11.9     &   M90\\
CGCG 049-057 *        & 119.1    $\pm$ 25.3        &      B08, P12                              &     ESO 244-G012         &  170.1  $\pm$ 51.0    &   A07\\
ESO 320-G030 *        &  225.3   $\pm$ 67.6        &      B08                                       &     ESO 255-IG007        &  89.1   $\pm$ 17.8     &   M90\\
IRASF 18293-3413 *  &  686.1   $\pm$ 205.8      &      B08                                       &     ESO 286-IG019        &  71.5   $\pm$  21.5    &   G99\\
Mrk 331 *                    &  371.2   $\pm$ 86.2       &      Y95, S91                              &      ESO 467-G027         & 191.7  $\pm$ 57.5    &    A07\\
NGC 3256 *                & 1222.8  $\pm$ 366.8      &      B08                                       &     ESO 507-G070         &  152.0  $\pm$ 45.6    &   P12\\
NGC 5135 *                &  380.4   $\pm$ 61.4        &      S91, B08, P12                      &     IC 4280                     &  310.5  $\pm$ 93.2     &   A07\\
NGC 6240 *               &  290.9   $\pm$ 44.3        &      Y95, S91, B08, P12              &     IC 4734                     &  367.2  $\pm$ 110.2   &   G93\\
NGC 7469 *                &  298.0   $\pm$  89.4       &      P12                                       &     IC 5298                     &  72.0    $\pm$ 14.4     &   S91\\ 
NGC 7552 *                &  652.0   $\pm$ 195.6      &      B08                                       &     IRASF 01417+1651  &  75.0   $\pm$  22.5     &   G99\\
NGC 7771 *                &  370.3   $\pm$ 84.2        &      Y95, S91                               &     IRASF 10565+2448  &  73.5   $\pm$  11.8     &   S91, B08, P12\\
NGC 6286 *                &  213.3   $\pm$  49.0       &      Y95, S91                               &     IRASF 16399-0937   & 118.1   $\pm$ 35.4     &   B08\\
CGCG 052-037 *        &  63.0     $\pm$ 18.9        &      P12                                       &     MCG-03-04-014        &  178.0  $\pm$ 53.4     &   P12\\
NGC 0828 *                &  389.3   $\pm$ 60.8        &      Y95, S91, B08, P12              &     NGC 0023                 &  203.0   $\pm$  37.1   &   S91, Y95, A07\\
NGC 2369 *                &  553.8   $\pm$ 166.4      &      B08                                       &     NGC 0317                 &  180.6   $\pm$ 72.4     &   Z99\\
NGC 6701 *                &  227.2   $\pm$ 42.5        &      Y95, P12, S91                       &     NGC 0695                 &  190.0  $\pm$ 43.8     &   Y95,S91 \\
UGC 02238 *              &  180.0   $\pm$  36.0       &      S91                                        &    NGC 1275                 &  180.0   $\pm$ 72.0    &    L89 \\
VV 340 *                     &  426.6   $\pm$ 128.0      &      G99                                       &     NGC 3110                 &   390.0  $\pm$ 78.0    &    S91\\
NGC 2623 *                &  155.8   $\pm$  28.6       &      S91, Y95, P12                       &     NGC 4194                &   129.9  $\pm$  24.5    &   S91, A07, Y95\\
MCG+12-02-001 *      &  230.0   $\pm$  46.0       &      G12                                       &     NGC 5104                 &   158.7  $\pm$  33.7    &   A07, P12\\
IC 1623 *                    &  557.0   $\pm$  111.9      &      S91, P12                                &    NGC 5653                 &   173.0  $\pm$  32.2    &   S91, P12\\
NGC 0232 *                &  310.5  $\pm$  93.2        &      A95                                        &    NGC 5936                 &   180.4   $\pm$ 41.6    &   Y95,S91 \\
IRASF 05189-2524   &  72.8    $\pm$  24.8        &      S91, B08, P12                        &     NGC 5990                &   299.8   $\pm$ 78.2    &   A07, P12\\
IRASF 17207-0014   &  160.0  $\pm$  48.0        &      P12                                         &     NGC 6621                &   159.5  $\pm$  41.1    &   S91, Z99\\
Mrk 231                     &  103.6  $\pm$  37.4        &     Y95, B08, A07, P12                 &     NGC 7591                 &    171.7  $\pm$ 68.7    &   L98\\
Mrk 273                     &   82.8   $\pm$  14.5        &     S91, B08, P12                         &     NGC 7592                &   177.0   $\pm$ 35.4    &   S91\\
NGC 1614                 &  247.7  $\pm$ 37.8         &     Y95, B08, A07, S91                 &     NGC 7674                &   120.0   $\pm$  24.0   &   S91\\
NGC 7130                 &  326.7  $\pm$ 70.9        &     B08, A07                                  &     NGC 7679                &    252.3  $\pm$ 100.9  &   W89 \\
UGC 05101               &  71.7    $\pm$  15.2        &     B08, P12                                 &     UGC 02608               &   413.1   $\pm$ 123.9   &  A07\\
CGCG 448-020         &  100.7  $\pm$  27.8        &     B08, P12                                 &     UGC 02982               &   277.5   $\pm$  70.1    &  Y95, A07\\
UGC 03094               &  207.9  $\pm$  62.4        &     A07                                          &     IRAS 05442+1732    &  148.5    $\pm$  44.6    &  A07\\
UGC 03351               &  411.1  $\pm$ 123.3       &     B08                                          &     NGC 4418                 &  137.85  $\pm$  24.5     &  Y95, B08, S91, P12\\
UGC 08739               & 118.0   $\pm$  24.0        &     P12                                          &     NGC 6052                 &   132.8   $\pm$ 39.9     &  A07\\
UGC 11041               &  232.2  $\pm$  69.7        &     A07                                          &     UGC 02369               &   194.0   $\pm$  58.2     & G99\\
VV 705                      &  90.6    $\pm$  23.4        &     A07, P12\\
  \noalign{\smallskip}
 \hline
\end{tabular}

\tablecomments{$^{(a)}$ The symbols $\lq\lq$*" represent the 23 galaxies which have \Cone\ detections.}
\tablecomments{$^{(b)}$ Reference to CO$\,(1-0)$ data: B08 = Baan et al. 2008, P12 = Papadopoulos et al. 2012, Y95 = Young et al. 1995, S91 = Sanders et al. 1991, G99 = Gao $\&$ Solomon 1999, G12 = Garc\'{i}a-Burillo et al. 2012, A95 = Andreani et al. 1995, A07 = Albrecht et al. 2007, M90 = Mirabel et al. 1990, G93 = Garay et al. 1993, Z99 = Zhu et al. 1999, L98 = Lavezzi et al. 1998, W89 = Wiklind $\&$ Henkel 1989.}
\end{table*}

\begin{table*}
\centering
\begin{minipage}[]{120mm}
\caption{Physical parameters of the sample}\centering
\label{tab2}\end{minipage}
\setlength{\tabcolsep}{5.0pt}
\small
 \begin{tabular}{cccccccccccc}
  \hline\noalign{\smallskip}
   galaxy  &   $R_\mathrm{[CI]}$   & Tex  &  $M_\mathrm{[CI]} $ $^{a}$ &  $M\mathrm{(H_{2})}_\mathrm{{CO}}$  & $M\mathrm{(H_{2})}_\mathrm{{[CI](1-0)}}$  &$M\mathrm{(H_{2})}_\mathrm{{[CI](2-1)}}$ \\
              &                                &  (K)   & $(10^6 \mathrm{M_\odot})$ & $(10^9 \mathrm{M_\odot}) $  &  $(10^9 M_\odot) $ &$(10^9 M_\odot) $& \\
              
  \hline\noalign{\smallskip}

Arp 193                     &  0.54  $\pm$ 0.07      &   28.4  $\pm$ 2.9  &   1.5 $\pm$  0.2    & 4.3 $\pm$ 0.9 &    8.3   $\pm$  1.1   &   10.5  $\pm$  2.2 \\
Arp 220                     &  0.35  $\pm$ 0.07      &   21.7  $\pm$ 2.3  &   2.5 $\pm$  0.5    & 6.6 $\pm$ 1.3 &   14.2  $\pm$  2.6   &   17.8  $\pm$  4.0 \\
CGCG 049-057         &  0.25  $\pm$ 0.05    &  18.2   $\pm$ 1.6  &   0.5 $\pm$  0.1    & 1.0 $\pm$ 0.2 &    2.9   $\pm$  0.6   &   3.6    $\pm$  0.9 \\
ESO 320-G030         &  0.34  $\pm$ 0.05    &   21.4  $\pm$ 1.8  &   0.3 $\pm$  0.05  &  0.8 $\pm$ 0.2 &   1.7   $\pm$  0.3   &    2.1   $\pm$  0.5 \\
IRASF 18293-3413   &  0.34  $\pm$ 0.03    &   21.3  $\pm$ 1.2  &   3.5 $\pm$  0.4    &  9.8 $\pm$ 3.0 &   19.5 $\pm$ 2.0    &    24.4 $\pm$  5.5 \\
Mrk 331                     &  0.38  $\pm$ 0.04    &   22.6  $\pm$ 1.5  &   1.0 $\pm$  0.1  &  4.5 $\pm$ 1.1 &   5.4   $\pm$  0.6   &    6.8   $\pm$  1.5  \\
NGC 3256                 &  0.45  $\pm$ 0.04    &   25.3  $\pm$ 1.6  &   1.1 $\pm$  0.1  &  3.6 $\pm$ 1.1 &   5.9   $\pm$  0.5   &    7.5   $\pm$  1.6  \\
NGC 5135                 &  0.43  $\pm$ 0.04    &   24.4  $\pm$ 1.5  &   1.2 $\pm$  0.1  &  2.8 $\pm$ 0.4 &   6.8   $\pm$  0.6   &   8.5  $\pm$  1.9 \\
NGC 6240                 &  0.81  $\pm$ 0.09    &   40.7  $\pm$ 4.8  &   3.5 $\pm$  0.3    &  7.5 $\pm$ 1.2 &   19.6 $\pm$  1.8   &    25.2 $\pm$  4.6 \\
NGC 7469                 &  0.44  $\pm$ 0.05    &   24.9  $\pm$ 1.7  &   1.1 $\pm$  0.1  &  2.9 $\pm$ 0.9 &   5.9   $\pm$  0.6   &    7.5   $\pm$  1.6  \\
NGC 7552                 &  0.46  $\pm$ 0.05    &   25.5  $\pm$ 1.7  &   0.3 $\pm$  0.03  &  0.7 $\pm$ 0.2 &   1.7   $\pm$  0.2   &    2.2   $\pm$  0.5 \\
NGC 7771                 &  0.31  $\pm$ 0.03    &   20.1  $\pm$ 1.2  &   1.1 $\pm$   0.1   &  2.7 $\pm$ 0.6 &   5.9   $\pm$  0.7   &    7.4   $\pm$  1.7 \\
NGC 6286                 &  0.20    $\pm$ 0.03    &   16.5  $\pm$ 1.1  &   2.5 $\pm$  0.4    &  3.0 $\pm$ 0.7 &  14.0  $\pm$  2.2   &    17.4 $\pm$  4.3 \\
CGCG 052-037         &  0.23  $\pm$ 0.05    &   17.7  $\pm$ 1.5  &   1.8 $\pm$  0.3    &  1.6 $\pm$ 0.5 &  9.8  $\pm$  2.0   &    12.3 $\pm$  3.0 \\
NGC 0828                 &  0.23  $\pm$ 0.05    &   17.2  $\pm$ 1.7  &   1.7 $\pm$  0.4    &  4.4 $\pm$ 0.7 &  9.3  $\pm$  2.1   &    11.6 $\pm$  2.9 \\
NGC 2369                 &  0.33  $\pm$ 0.05    &   20.9  $\pm$ 1.7  &   0.8 $\pm$  0.1    &  2.5 $\pm$ 0.7 &   4.2   $\pm$  0.7   &    5.3   $\pm$  1.2 \\
NGC 6701                 &  0.18  $\pm$ 0.03    &   15.6  $\pm$ 1.1  &   1.6 $\pm$  0.3    &  1.8 $\pm$ 0.3 &   9.1   $\pm$  1.6   &    11.3 $\pm$  2.8 \\
UGC 02238               &  0.25  $\pm$ 0.06    &   18.4  $\pm$ 1.9  &   1.4 $\pm$  0.3    &  3.0 $\pm$ 0.6 &   8.0   $\pm$  1.8   &    10.0 $\pm$  2.4 \\
VV 340                      &  0.20    $\pm$ 0.03    &   16.5  $\pm$ 1.0  &   5.0 $\pm$  0.8    & 19.9 $\pm$ 6.0 &   27.9 $\pm$  4.3   &    34.7 $\pm$  8.5 \\
NGC 2623                 &  0.51    $\pm$ 0.09      &   27.5  $\pm$ 3.3  &   0.6 $\pm$  0.1    &  2.1 $\pm$ 0.4 &   3.4   $\pm$  0.6   &    4.3   $\pm$  0.9 \\
MCG+12-02-001       &  0.54  $\pm$ 0.09      &   28.4  $\pm$ 3.6  &   0.5 $\pm$ 0.08   &  2.2 $\pm$ 0.4 &   2.8   $\pm$  0.5   &    3.5   $\pm$  0.7 \\
IC 1623                     &  0.35  $\pm$ 0.04    &   21.6  $\pm$ 1.4  &   1.6 $\pm$  0.2    &  7.8 $\pm$ 1.6 &   9.0   $\pm$  1.0   &   11.3   $\pm$  2.5 \\
NGC 0232                 & 0.26  $\pm$ 0.04     &  18.5   $\pm$ 1.5  &   2.1 $\pm$  0.4   &   5.4 $\pm$ 1.6 &  11.7 $\pm$ 2.0    &   14.6  $\pm$   3.4 \\
ESO 264-G036         & 0.23  $\pm$ 0.05     &  17.5 $\pm$ 1.6     &   2.0 $\pm$ 0.4   &                           &   11.1 $\pm$ 2.2  &  13.8   $\pm$  3.3 \\
IRAS 12116-5615     & 0.38  $\pm$ 0.09     &   22.6 $\pm$ 3.1    &   1.5 $\pm$ 0.3   &                           &    8.4  $\pm$ 2.0   &  10.6  $\pm$  2.3\\
UGC 12150              &  0.21  $\pm$ 0.05     &  16.8 $\pm$ 1.7    &   1.6  $\pm$ 0.4   &                           &   8.6  $\pm$  2.0   &  10.8  $\pm$  2.6\\
IRAS 13120-5453     &  0.42  $\pm$ 0.06     &   24.1 $\pm$ 2.1   &   4.8  $\pm$ 0.7   &                           &   26.9 $\pm$ 3.7  &   33.9  $\pm$  7.2\\
ESO 173-G015         & 0.55   $\pm$ 0.07     &  28.9  $\pm$ 2.6   &   0.5  $\pm$  0.06  &                           &   3.0  $\pm$  0.3  &  3.9  $\pm$  0.8 \\
IC 4687                    &  0.34   $\pm$ 0.05     &  21.4  $\pm$ 1.7  &  0.7  $\pm$  0.1 &                           &   4.0  $\pm$  0.6   &  5.1  $\pm$ 1.1\\
NGC 0034                &  0.25   $\pm$ 0.04     &  18.1   $\pm$ 1.5  &  0.8  $\pm$  0.1 &                          &   4.2  $\pm$  0.8   &  5.2  $\pm$  1.3\\


  \noalign{\smallskip}
 \hline
\end{tabular}
\tablecomments{$^{(a)}$ The carbon mass $M_\mathrm{[CI]}=6\times M(\mathrm{H_2})X_{\rm [CI]}$, is calculated using the \Cone\, line.}
\tablecomments{Column 2 is the line ratio of \Ctwo\ to \Cone; column 3 is the excitation temperature derived with the \CI\ lines; column 4 is the carbon mass calculated from \Cone; column 5 is the $\mathrm{H_2}$ gas mass calculated from the CO$\,(1-0)$ line; columns 6 and 7 are the $\mathrm{H_2}$ gas mass calculated from the \Cone\ and \Ctwo\ lines.}
\end{table*}

\section{Summary}

In this letter, we present the relations of the \Cone, \Ctwo\ lines with the CO\,(1$-$0) line for a sample of 71 (U)LIRGs which were observed with the Herschel SPIRE/FTS. We investigate the dependence of $L'_\mathrm{[CI](2-1)}$/$L'_\mathrm{CO(1-0)}$, $ L'_\mathrm{[CI](1-0)}$/$L'_\mathrm{CO(1-0)}$ and the \CI\ intensity ratio $R_\mathrm{[CI]}$ on far-infrared color ($f_{60}/f_{100}$). We also calculate the conversion factors of $\alpha_\mathrm{[CI](1-0)}$ and $\alpha_\mathrm{[CI](2-1)}$, based on the assumption that the carbon is optically thin, and on the \CI$-$CO relation. Our main results are as follows:\\

1. There is an obvious correlation between $L'_\mathrm{[CI](1-0)}$ and $L'_\mathrm{CO(1-0)}$ in (U)LIRGs, i.e., $ \log L'_\mathrm{CO(1-0)} = (-0.23 \pm 1.32) + (1.10\pm 0.15) \log L'_\mathrm{[CI](1-0)} $. $L'_\mathrm{[CI](2-1)}$ also correlates well with $L'_\mathrm{CO(1-0)}$, namely, $ \log L'_\mathrm{CO(1-0)} =  1.46 \pm 0.69 + (0.97 \pm 0.08)\, \log L'_\mathrm{[CI](2-1)} $. These results imply that the \Cone, \Ctwo\ lines can be used as total molecular tracers at least for (U)LIRGs.\\

2. We find that $ L'_\mathrm{[CI](2-1)}$/$L'_\mathrm{CO(1-0)}$ depends weakly on $f_{60}/f_{100}$, while $ L'_\mathrm{[CI](1-0)}$/$L'_\mathrm{CO(1-0)}$ has no correlation with $f_{60}/f_{100}$.\\

3. Based on the \CI$-$CO relations, we derive the conversion factors of $\alpha_\mathrm{[CI](1-0),CO} = 3.6$ and $\alpha_\mathrm{[CI](2-1),CO} = 12.5$\,$\mathrm {(K\,km\,s^{-1}\,pc^2)^{-1}}$. Whereas the conversion factors derived from a more direct method are $\alpha_\mathrm{[CI](1-0)} = 7.6$ and $\alpha_\mathrm{[CI](2-1)} = 27.5$\,$\mathrm {(K\,km\,s^{-1}\,pc^2)^{-1}}$ by assuming a constant \CI\ abundance. The accuracy is about a factor of $2-3$.\\

The authors thank the referee for useful suggestions.
This work is partially supported by the Natural Science Foundation of China under grant NOs. 11673057, U1531246, 11420101002, 11673028 and 11311130491. YG thanks partial support of the CAS Key Research Program of Frontier Sciences. Z-Y.Z. acknowledges support from ERC in the form of the Advanced Investigator Programme, 321302, COSMICISM. 



\end{document}